\documentclass[twocolumn]{aastex63}

\usepackage{xspace}
\usepackage{booktabs}
\usepackage{multirow}
\usepackage{layouts}
\usepackage{amsmath}

\newcommand{\Vshell}[2]{\mathbf{#1}^\mathrm{#2}}

\newcommand{\Va}{\mathbf{v}_{\mathrm{A}}}

\newcommand{\V}[1]{\mathbf{#1}}
\newcommand{\bra}[1]{\left ( #1 \right )}
\newcommand{\ang}[1]{\left < #1 \right >}

\newcommand{\Ms}{\mathrm{M_s}}
\newcommand{\Ma}{\mathrm{M_a}}

\newcommand{\pb}{\beta_{\mathrm{p}}}

\renewcommand{\Re}{\mathrm{Re}}
\newcommand{\Rm}{\mathrm{Rm}}
\newcommand{\Pm}{\mathrm{Pm}}
\newcommand{\pth}{p_\mathrm{th}}
\newcommand{\keq}{k_\mathrm{eq}}

\newcommand{\Ekin}{E_\mathrm{kin}}
\newcommand{\Emag}{E_\mathrm{mag}}

\newcommand{\T}[1]{\mathcal{T}_\mathrm{#1}}

\newcommand{\kathena}{\texttt{K-Athena}\xspace}
\newcommand{\kokkos}{\texttt{Kokkos}\xspace}
\newcommand{\athenapp}{\texttt{Athena++}\xspace}

\begin{document}
\title{As a matter of tension  -- kinetic energy spectra in
MHD turbulence}
\shorttitle{Kinetic energy spectra in MHD turbulence}
\shortauthors{Grete, O'Shea \& Beckwith}
\correspondingauthor{Philipp Grete}
\email{grete@pa.msu.edu.}

\author[0000-0003-3555-9886]{Philipp Grete}
\affiliation{
Department of Physics and Astronomy,
Michigan State University, East Lansing, MI 48824, USA}
\author[0000-0002-2786-0348]{Brian W. O'Shea}%
\affiliation{
Department of Physics and Astronomy,
Michigan State University, East Lansing, MI 48824, USA}
\affiliation{
Department of Computational Mathematics, Science and Engineering,
Michigan State University, East Lansing, MI 48824, USA}
\affiliation{
National Superconducting Cyclotron Laboratory,
Michigan State University, East Lansing, MI 48824, USA}

\author[0000-0002-5610-8331]{Kris Beckwith}
\affiliation{%
Sandia National Laboratories, Albuquerque, NM 87185-1189, USA
}


\begin{abstract}

Magnetized turbulence is ubiquitous in many astrophysical and terrestrial systems but
no complete, uncontested theory even in the simplest form, magnetohydrodynamics (MHD), exists.
Many theories and phenomenologies focus on the joint (kinetic and magnetic) energy fluxes
and spectra.
We highlight the importance of treating kinetic and magnetic energies separately to shed
light on MHD turbulence dynamics.
We conduct an implicit large eddy simulation of subsonic, super-Alfv\'enic MHD turbulence 
and analyze the scale-wise energy transfer over time.
Our key finding is that the kinetic energy spectrum develops a scaling of approximately
$k^{-4/3}$ in the stationary regime as the kinetic energy cascade is suppressed by magnetic
tension.
This motivates a reevaluation of existing MHD turbulence theories with respect to a
more differentiated modeling of the energy fluxes.

\end{abstract}

\section{Introduction}
While our understanding of incompressible hydrodynamic turbulence has significantly advanced
over the past decades, many critical questions in the realm of compressible 
magnetohydrodynamic (MHD) turbulence remain unanswered.
This regime is of particular interest in both astrophysics and in terrestrial systems where
processes on a huge variety of scales are either governed or at least influenced
by MHD turbulence.
Astrophysical examples include energy transport in the solar convection zone
\citep{Canuto1998,Miesch2005}, angular momentum transport and energy release
in accretion disks \citep{Balbus1998}, the core collapse supernova mechanism
\citep{Couch:2015,Mosta:2015},
the interstellar medium with its star-forming molecular clouds
\citep{Vazquez-Semadeni2015,Falgarone2015,Klessen2016}, and clusters of
galaxies that can be used to determine cosmological parameters
\citep{Brunetti2015,Bruggen2015}. In the terrestrial context, this is
of interest for a range of plasma experiments, such as laser-produced colliding plasma flows, Z-pinches, 
and tokamaks
\citep[see, e.g.,][]{Tzeferacos2017,haines_zpinch,2009NucFu..49e5001M,2013NucFu..53h3007R}.

At the same time, MHD turbulence theory and phenomenology also made significant progress
from early isotropic models \citep{Iroshnikov1964,Kraichnan1965}, to
critically balanced turbulence \citep{Sridhar1994,Goldreich1995}, to
dynamic alignment \citep{Boldyrev2009}, but 
it is still a highly debated topic -- see, e.g., \citet{Galtier2016,Beresnyak2019}
for recent reviews.
Different theories make a variety of predictions for the scaling of the energy spectra
depending on the strength of the mean magnetic magnetic field (either external or local), on the
cross helicity (balanced versus unbalanced turbulence), and on the magnetic helicity 
(encoding the topology of the magnetic field configuration).
In the majority of cases scaling predictions are only concerned with the
total energy spectrum ($E(k) = \Ekin (k) + \Emag (k)$ with wavenumber $k$)
and assume a moderate or strong background field so
that dynamics are differentiated between parallel and perpendicular to the
mean field. Thus, there is no differentiation between the kinetic ($\Ekin (k)$)
and magnetic ($\Emag (k)$) energy spectra.
A complementary theoretical approach to modeling magnetohydrodynamic turbulence is the 
use of shell models, which are a computationally-inexpensive semi-analytical means of 
modeling turbulence.
Notable examples of this include \citet{Biskamp1994,Frick1998,Plunian2007}
who also observe, for example, flatter spectra, spectral breaks, and different
scaling behavior of the kinetic and magnetic energy spectra.
However, the behavior strongly depends on the characteristics of the system being modeled 
(with, e.g., properties of the system such as a mean magnetic field, helicity and cross helicity contributing
significantly to the observed outcomes similarly to the locality of the interactions considered). 
By contrast to predictions from analytic and semi-analytic modeling efforts, numerous computational studies of magnetized turbulence have reported \emph{different} scaling behavior of kinetic and magnetic energy spectra
\citep{ Haugen2004,Moll2011,Teaca2011,Eyink2013,Porter2015,Grete2017a,Bian2019}
and, perhaps even more importantly, different scaling behavior of kinetic and magnetic energy spectra
has been reported in observations of the solar wind  \citep{Boldyrev2011}.

In order to gain a deeper insight into this discrepancy, we present and analyze the evolution and stationary state of the
kinetic and magnetic energy spectra and fluxes separately in the
context of an implicit large eddy simulation of ideal MHD turbulence in its simplest
configuration (vanishing mean field, cross-helicity, and magnetic helicity).
We confirm prior results \citep{ Haugen2004,Moll2011,Teaca2011,Eyink2013,Porter2015,Grete2017a,Bian2019}
that the kinetic and magnetic energy spectra exhibit different
scaling behavior. In particular, we find that the kinetic energy spectrum exhibits a scaling close
to $k^{-4/3}$ -- i.e., it is shallower than the spectra predicted in the theories above, which mostly 
range between $k^{-3/2}$ and $k^{-5/3}$. We further demonstrate, 
using a shell-to-shell energy transfer analysis, that this ``shallow'' kinetic energy spectrum
is associated with magnetic tension, which acts to suppress the kinetic energy cascade
and provides the major contribution in the energy flux from large to small scales. This
result is in marked contrast with incompressible \emph{hydrodynamic} turbulence, where the kinetic energy
cascade is the \emph{only} means of transferring energy between scales in a self-similar fashion
(which in turn leads to the emergence of the $k^{-5/3}$ scaling) and departures from this expected
scaling in hydrodynamic turbulence simulations and experiments have been associated with the existence of
``bottlenecks'' \citep{Falkovich1994,Schmidt2006353, Frisch2008,Donzis2010,Kuechler2019,Agrawal2020}. As such, the results presented in this work demonstrate
the rich physics phenomenology that can operate even in the simplest scenarios (vanishing mean field, cross-helicity, and magnetic helicity) where magnetic tension is dynamically important and further serve to highlight the necessary
ingredients that MHD turbulence theory and phenomenology should incorporate in order to explain scalings of kinetic and magnetic energy observed in both simulation \emph{and} observation of magnetized turbulence.

The rest of this paper is structured as follows.
In Section~\ref{sec:method} we introduce the simulation setup and
summarize the energy transfer analysis. In Section~\ref{sec:results}
we present the kinetic and magnetic energy spectra,
their temporal evolution, and scale dependent energy dynamics.
Finally, in Section~\ref{sec:summary}, we summarize our results,
the limitations of the simulations upon which they are based and
discuss the implications for both modeling of magnetohydrodynamic
turbulence and astrophysical systems.

\section{Method}
\label{sec:method}

\subsection{Simulation setup}

We use the open source code, \kathena \citep{kathena},
a performance portable implementation of \athenapp \citep{Stone2019} based on
\kokkos \citep{Edwards2014}, to solve the ideal MHD equations\footnote{
\kathena is available at \url{https://gitlab.com/pgrete/kathena}. Commit \texttt{e5faee49}
was used to run the simulation and the parameter file (\texttt{athinput.fmturb})
is contained in the supplemental material for this paper.
}.
The second-order finite volume scheme employed is comprised of a Van-Leer integrator,
constrained transport MHD algorithm, piecewise-linear reconstruction, and Roe Riemann solver, 
\citep[see][for more details on the numerical method]{Stone2009}.
Given that no explicit physical dissipative terms are present dissipation is
purely numerical; as such the simulations presented here utilize the implicit large eddy simulation (ILES) technique \citep{grinstein2007implicit}. Turbulent driving is accomplished through a stochastic forcing approach described by \citep{Schmidt2009}, implemented within \kathena using a communication-avoiding algorithm for efficient large scale parallel simulations on GPUs.

We conduct a single simulation of a cubic domain with side length of 1 (if not noted otherwise
all units are in code units) and 
periodic boundary conditions  on a $2{,}048^3$ grid.
The plasma is initially at rest (velocity $\V{u} = 0$) with uniform density ($\rho = 1$)
and thermal pressure ($\pth = 1$).
The initial magnetic field configuration ($\V{B}_0 = \nabla \times \V{A}_0$ 
with $\V{A}_0 = (0,0,r_0 - r)$ for $r < r_0$ with $r = \sqrt{\bra{x-0.5}^2 + \bra{y-0.5}^2}$)
is a cylinder in the z-direction with radius $r_0 = 0.4$
and centered in the xy-plane, i.e., there is no magnetic flux going through any of the outer
surfaces. 
The initial magnetic field strength is comparatively weak with $\ang{\Emag} = 0.00125$ corresponding
to a plasma beta (ratio of thermal to magnetic pressure) of $\pb = 800$.
The plasma is kept approximately isothermal using an adiabatic equation of state with
adiabatic index of $\gamma = 1.0001$.

In order to reach a state of stationary turbulence, we employ a large scale mechanical driving force
(having an inverse parabolic shape with a peak at $k = 2$, where $k$ is the normalized wavenumber).
The driving field is purely solenoidal and has an autocorrelation time of 1.0
so that no artificial compressible modes are injected \citep{Grete2018a}.
In the stationary regime the integral length is 
$L = \int \Ekin(k)/k \;\mathrm{d}k/\int \Ekin(k) \mathrm{d}k = 0.32$ 
(i.e., slightly smaller than the forcing scale at 0.5), the root mean square (RMS) sonic Mach number is $\Ms = 0.54$, 
the resulting large eddy turnover time is $T = L /(\Ms c_s) = 0.59$,
the RMS Alfv\`enic Mach number is $\Ma = 2.8$, and the mean plasma beta is $\pb = 54$.

\subsection{Energy transfer analysis}
For a detailed analysis of the energy dynamics we apply the shell-to-shell energy transfer
analysis presented in \citet{Grete2017a}, which is an extension of \citet{Alexakis2005} to
the compressible regime.
The key idea is to separate energy transfers by their source (some
energy budget at some spatial scale $Q$), sink 
(some budget at some scale $K$), and a mediator.
Given the isothermal nature of the simulation, we focus on the kinetic and magnetic energy budget only
and neglect a detailed analysis of the internal energy budget \citep[cf.][]{Schmidt2019} or
non-isothermal statistics \citep{Grete2020}.

In general, the energy transfers are given by
\begin{align}
\T{XY} (Q,K) \quad \text{with} \quad \mathrm{X,Y} \in \{\mathrm{U,B} \}
\end{align}
expressing energy transfer (for $\T{} > 0$) from shell $\mathrm{Q}$ of energy budget $\mathrm{X}$
to shell $\mathrm{K}$ of energy budget $\mathrm{Y}$.   $\mathrm{U}$
and $\mathrm{B}$ represent the kinetic and magnetic energy budgets, respectively.
More specifically, the energy transfers are
\begin{align}
\T{UU}(Q,K) = - \int \Vshell{w}{K} \cdot \bra{\V{u} \cdot \nabla} \Vshell{w}{Q}  +
\frac{1}{2} \Vshell{w}{K} \cdot \Vshell{w}{Q} \nabla \cdot \V{u} \mathrm{d}\V{x} \\
\T{BB}(Q,K) = - \int \Vshell{B}{K} \cdot \bra{\V{u} \cdot \nabla} \Vshell{B}{Q}  +
\frac{1}{2} \Vshell{B}{K} \cdot \Vshell{B}{Q} \nabla \cdot \V{u} \mathrm{d}\V{x} 
\end{align}
for kinetic-to-kinetic (and magnetic-to-magnetic) transfers via advection and compression,
\begin{align}
\T{BUT}(Q,K) = \int \Vshell{w}{K} \cdot  \bra{ \Va \cdot \nabla} \Vshell{B}{Q}  \mathrm{d}\V{x} \\
\T{UBT}(Q,K) =  \int  \Vshell{B}{K} \cdot \nabla \cdot \bra{\Va \otimes \Vshell{w}{Q}} \mathrm{d}\V{x}
\end{align}
for magnetic-to-kinetic (and kinetic-to-magnetic) energy transfer via magnetic tension, and
\begin{align}
\T{BUP}(Q,K) = - \int \frac{\Vshell{w}{K}}{2 \sqrt{\rho}} \cdot \nabla \bra{ \V{B} \cdot \Vshell{B}{Q}}  \mathrm{d}\V{x}\\ 
\T{UBP}(Q,K) = - \int \Vshell{B}{K} \cdot \V{B}  \nabla \cdot \bra{\frac{\Vshell{w}{Q}}{2 \sqrt{\rho}}}  \mathrm{d}\V{x} 
\end{align}
for magnetic-to-kinetic (and kinetic-to-magnetic) energy transfer via magnetic pressure.
Here, $\V{w} = \sqrt{\rho} \V{u}$ is a mass-weighted velocity chosen so that the spectral kinetic energy 
density based on $\frac{1}{2}w^2$ is a positive definite quantity \citep{Kida1990} and $\Va$ is the Alfv\'en velocity.

The velocity $\Vshell{w}{K}$ and magnetic field $\Vshell{B}{K}$ in a shell K (or Q) are obtained by a sharp spectral
filter in Fourier space with logarithmic spacing.
The bounds are given by $1$ and $2^{n/4 + 2}$ for $n \in \{ -1,0,1,\ldots,36\}$.
Shells (uppercase, e.g., K) and wavenumbers (lowercase, e.g., $k$) obey a direct mapping, i.e.,
$K=24$ corresponds to $k \in (22.6,26.9]$.

Given the low sonic Mach number of the simulation (i.e., limited density variations) differences
between the shell filtered transfers and transfers obtained through our coarse-graining approach (similar
to the formalism employed in large eddy simulations) are expected to be negligible \citep{Aluie2013,Yang2016}.

\section{Results}
\label{sec:results}

\subsection{Emergence of a power law in $\Ekin (k)$}

\begin{figure}[htbp]
\centering
\includegraphics{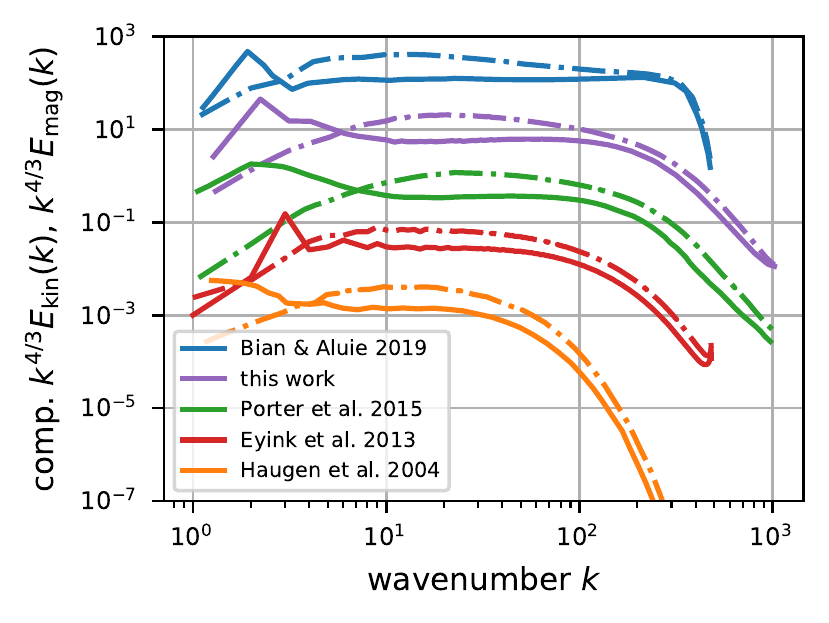}
\caption{
Kinetic (solid) and magnetic (dashed) energy spectra
reported in literature from simulations with various numerical
schemes, compensated by $k^{4/3}$:
pseudo-spectral DNS of incompressible MHD with hyperdissipation\citep[Fig.~10]{Bian2019},
ILES of compressible, ideal MHD \citep[Fig.~3]{Porter2015} similar to this work,
pseudo-spectral DNS of incompressible MHD \citep[Fig.~2]{Eyink2013}, and
higher-order finite difference DNS of compressible MHD with hyperdissipation \citep[Fig.~7]{Haugen2004}.
All spectra have in common that magnetic energy dominates all scales 
smaller than the forcing scale and that the kinetic energy spectrum exhibits
a region with scaling close to $k^{4/3}$.
(Lines are vertically offset for increased readability.)
}
\label{fig:slope-comp}
\end{figure}

In MHD turbulence simulations (independent of numerical method such as pseudospectral DNS,
higher-order finite difference, or finite volume ILES)
without a strong mean field ($B_0 \ll \ang{u}_\mathrm{RMS}$)
and magnetic Prandtl number $\Pm \approx 1$ two important features emerge in the
kinetic and magnetic energy spectra when plotted separately, see Figure~\ref{fig:slope-comp}
for a comparison
\citep{ Haugen2004,Moll2011,Teaca2011,Eyink2013,Porter2015,Grete2017a,Bian2019}.
First, the turbulent dynamo amplifies magnetic fields on all scales, resulting in $\Emag(k) > \Ekin(k)$
on all scales smaller than the forcing scales. 
Second, the kinetic energy spectrum develops a power law regime on the magnetically
dominated scales with a slope close to -4/3, i.e., shallower than the Kolmogorov slope of -5/3.

In order to understand the emergence of a flatter-than-Kolmogorov slope,
indicative of a less efficient energy cascade, we present a single simulation
in more detail in the following sections.

\subsection{Time evolution of the energy power spectra}

\begin{figure}[htbp]
\centering
\includegraphics{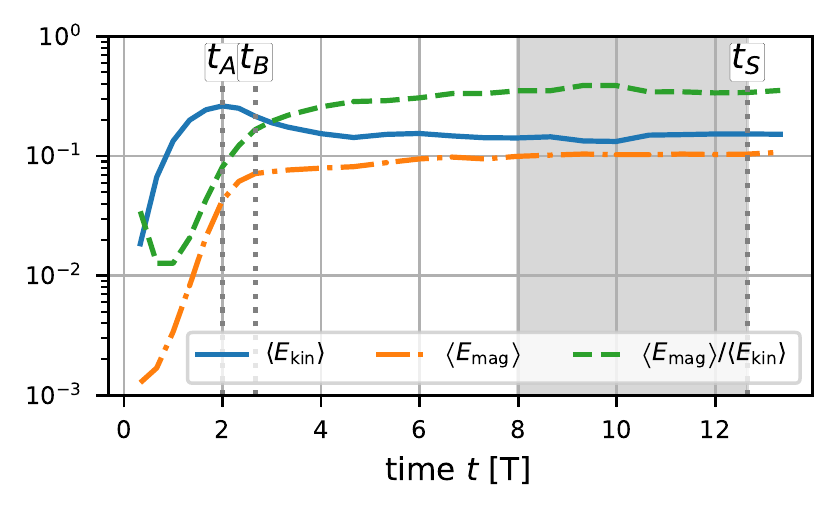}
\caption{Temporal evolution of
mean magnetic energy (orange dash-dotted line),
mean kinetic energy (blue solid),
and their ratio (green dashed).
The shaded gray area indicates the stationary regime.
Specific times $t_A$ (peak kinetic energy), $t_B$ (nonlinear dynamo), 
and $t_S$ (stationary)  correspond to snapshots that are analyzed in more detail.
}
\label{fig:en-evol}
\end{figure}

\begin{figure}[htbp]
\centering
\includegraphics{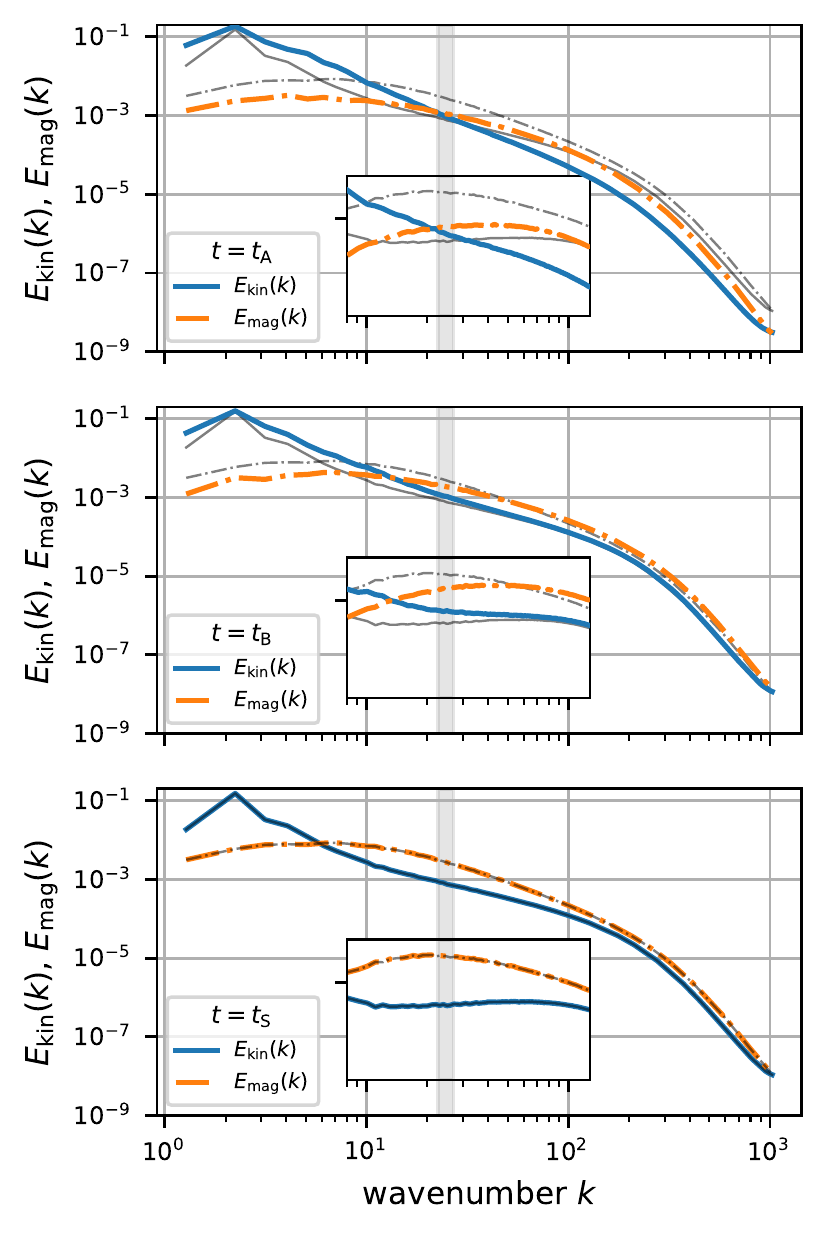}
\caption{Kinetic (blue solid) and magnetic (orange dash-dotted) energy spectra at different times.
The inset shows $8<k<64$ compensated by $k^{4/3}$ and
illustrates the flattening of the kinetic energy spectrum.
The thin lines in each panel illustrate the stationary state (bottom panel) for reference.
The gray area at $22.6<k \leq 26.9$ ($\widehat{=} K = 24$) indicates the scale that is 
used in the more detailed energy 
transfer analysis in Section~\ref{sec:en-dyn}.
 }
\label{fig:spectra-over-time}
\end{figure}

Figure~\ref{fig:en-evol} illustrates the evolution of the mean magnetic and kinetic energies
and their ratio over time.
First, the mean kinetic energy reaches its  peak value at time $t_A$.
The corresponding spectra\footnote{
A movie of the temporal evolution of the energy spectra (\texttt{Grete\_et\_al-spectra\_evol.mp4})
is available in the supplemental material.}
(top panel in Figure~\ref{fig:spectra-over-time}) show that the kinetic 
energy on small scales ($k \gtrsim 32$)
is lower than the stationary value (indicated by the thin black lines), whereas the kinetic energy on 
large scales is above the stationary value.
The magnetic energy spectrum crosses the kinetic energy spectrum at $\keq\approx24$ 
(where $\Ekin (k_{eq}) \approx \Emag (k_{eq})$)
so that the magnetic field become dynamically relevant on small scales.

At time $t_B$, which corresponds to the nonlinear phase of the dynamo, the kinetic
energy on small scales $k \gtrsim 50$ has reached its stationary value
(see center panel in
Figure~\ref{fig:spectra-over-time}).
Moreover, the kinetic energy spectrum shows a first indication of a spectral
break around $k\approx 24$ with steeper slope on large scales and a shallower 
slope on small scales.
The crossover of magnetic and kinetic energy has shifted towards larger scales and now
occurs around $\keq \approx 16$.

Finally, the stationary regime is reached after after $\approx 8T$ with $\Emag$
saturating at $\approx 0.28 \Ekin$.
In the stationary regime (represented by $t = t_S$) the crossover has shifted
to the largest scales $\keq \approx 8$ -- see bottom panel of Figure~\ref{fig:spectra-over-time}.
Further growth is inhibited due to the large scale purely mechanical force and
the magnetic energy is now dominant on all but the largest scales.
As a result the kinetic energy spectrum has been significantly flattened and now
exhibits a limited range $16 \lesssim k \lesssim 64$ with a shallower-than-Kolmogorov 
slope close to $\approx -4/3$.

In the following, we demonstrate how magnetic tension is responsible for
this flattening of the kinetic energy spectrum by suppressing the
kinetic energy cascade.

\subsection{Energy dynamics}
\label{sec:en-dyn}
In absence of explicit dissipation (and, thus, the explicit mean dissipation rate), 
all energy transfer rates are normalized using the mean total cross-scale flux at $k=26.9$
in the stationary regime as a proxy.
This choice has no influence on the actual results, but allows for an easier
comparison of relative magnitudes and with other results reported in the literature.

\subsubsection{Magnetic tension}
\label{sec:en-dyn-tension}
\begin{figure}[htbp]
\centering
\includegraphics{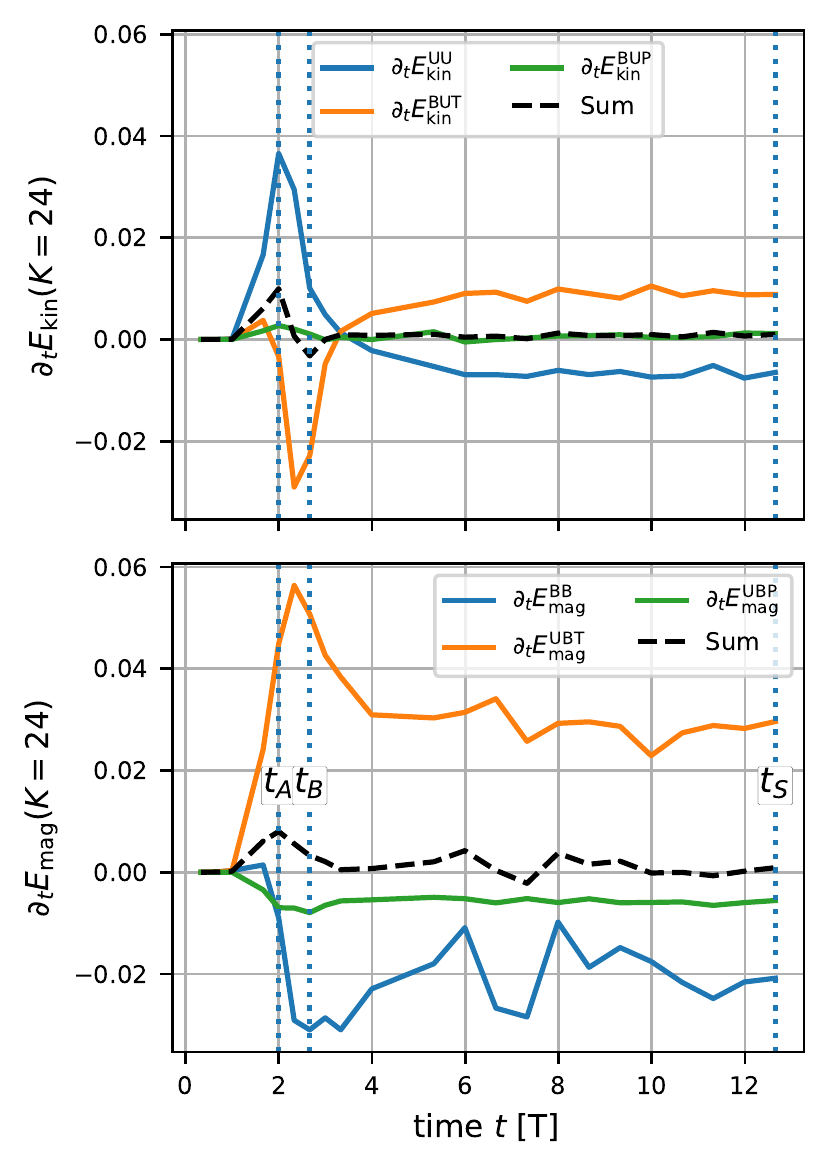}
\caption{Net rate of change in kinetic (top) and magnetic (bottom) energy at
$k=24$ over time.
Blue lines indicate energy transfer through advection, orange through 
magnetic tension, and green through magnetic pressures.
The pressure dilatation and forcing term are not shown as their contribution
is negligible.
}
\label{fig:total}
\end{figure}

The role of magnetic tension in shaping the kinetic energy spectrum becomes
apparent in Figure~\ref{fig:total}.
It shows the net rate of change in kinetic
energy (top row) and magnetic energy (bottom) row for the different mediators
over time and for the reference shell $K = 24$, i.e.,
\begin{align}
\partial_t \Ekin^\mathrm{XY}(K) &= \sum_Q \T{XY}(Q,K) \;\mathrm{and}\\
\partial_t \Emag^\mathrm{XY}(K) &= \sum_Q \T{XY}(Q,K) 
\end{align}
with XY $\in$ \{UU,BUT,BUP\} for the kinetic energy and 
XY $\in$ \{BB,UBT,UBP\} for the magnetic energy.
In other words, this is the net rate of change in energy at some scale K from
all other scales Q.

While at time $t_A$ the kinetic cascade ($\T{UU}$) is still contributing to a net
increase of kinetic energy on those scale, the rate of change by magnetic tension $\T{BUT}$
is negative, i.e., removing kinetic energy from $K = 24$.
The net effect remains positive.
At $t_B$ the dynamics have changed.
The kinetic cascade still contributes with a growth in kinetic energy, but magnetic
tension now dominates so that the net effect is a removal of kinetic energy from those scales.
This transfer of energy from the kinetic to the magnetic budget through magnetic 
tensions causes the flattening of the kinetic energy spectrum.

In the stationary regime the net rate of change in both kinetic and magnetic energy
fluctuates around 0 (otherwise the regime should not be considered stationary).
This balance is only maintained through energy transfers between kinetic and
magnetic energy budgets.
On average, the kinetic and magnetic cascades remove energy from intermediate scales of
their respective budgets (blue lines are negative) and this deficit is filled through transfers
mediated by magnetic tension between budgets (orange lines).

\begin{figure}[htbp]
\centering
\includegraphics{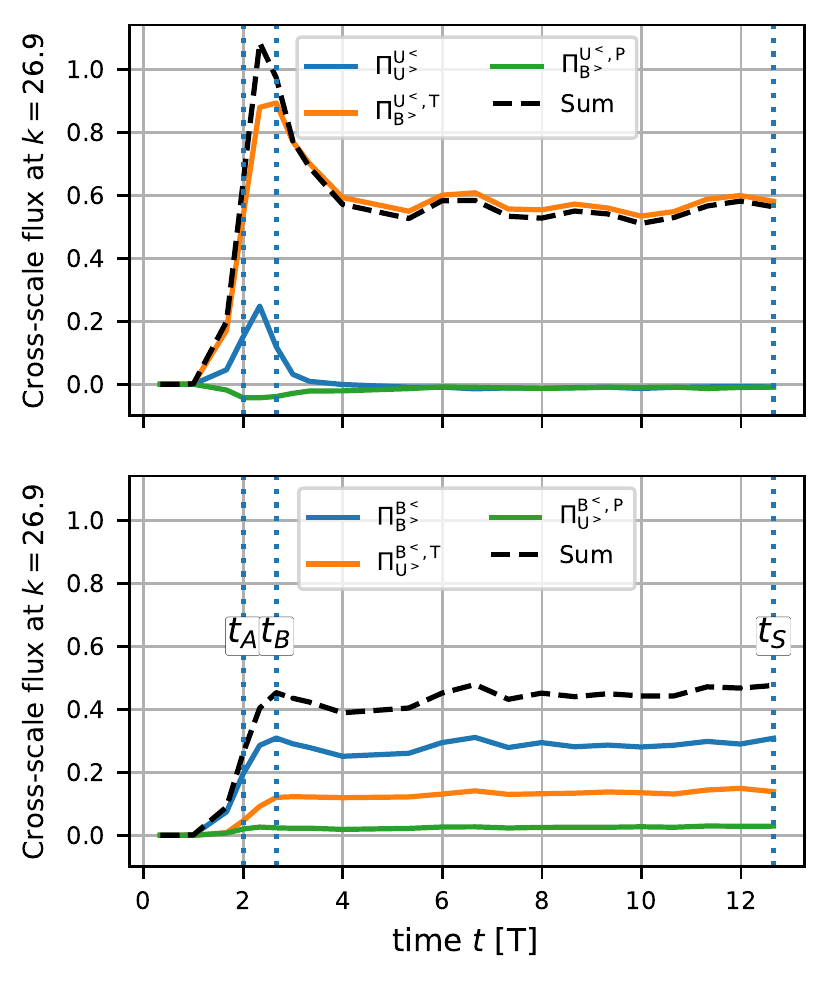}
\caption{Cross-scale energy transfer across $k=26.9$ over time, i.e., energy from all
budgets going from all larger scales ($k<26.9$) to the small scale ($k>26.9$) kinetic 
budget (top) and magnetic budget (bottom), respectively.
}
\label{fig:cross}
\end{figure}
The importance of magnetic tension is similarly observed in the cross-scale energy fluxes.
These fluxes are obtained from the individual transport terms via
\begin{align}
\Pi^\mathrm{U^<}_\mathrm{U^>}(k) &= \sum_{Q \leq k} \sum_{K>k} \T{UU}(Q,K)\;,  \\
\Pi^\mathrm{U^<,T}_\mathrm{U^>}(k) &= \sum_{Q \leq k} \sum_{K>k} \T{UBT}(Q,K)\;,  \\
\Pi^\mathrm{U^<,P}_\mathrm{U^>}(k) &= \sum_{Q \leq k} \sum_{K>k} \T{UBP}(Q,K)\;,  \\
\end{align}
for energy being transferred from the kinetic energy on all scales $\leq k$ to
to the kinetic and magnetic energies on scales smaller than $k$ by advection,
magnetic tension, and magnetic pressure, respectively.
The same notation applies to transfers from the large scale magnetic energy with U and B
indices exchanged.

Figure~\ref{fig:cross} illustrates the energy flux across $k=26.9$ over time 
from the large scale kinetic energy (top panel) and large scale magnetic energy
(bottom panel).
Again, the cross-scale flux initially increases in intensity for both
of the advection-related
transfers (blue lines).
While it peaks for $\Pi^\mathrm{B^<}_\mathrm{B^>}$ at $t_B$ and then remains at a constant
value, it already peaks for $\Pi^\mathrm{U^<}_\mathrm{U^>}$  at $t=t_A$ and afterwards
declines again to 0.
Transfers via magnetic tension (orange lines) from both large kinetic and magnetic scales
steadily growth till $t=t_B$.
Similar to the advection transfers, $\Pi^\mathrm{B^<, T}_\mathrm{U^>}$ remains constant after
the peak whereas $\Pi^\mathrm{U^<, T}_\mathrm{B^>}$ declines with the key difference
that the decline is not to 0 but to a non-zero value.
Moreover, it is the only remaining contribution for the kinetic energy cross-scale flux
(at that scale) and, overall, the dominating cross-scale flux is marginally ($\approx$15-20\%)
stronger than the combined fluxes from the large scale magnetic energy budget by advection
and tension.
In other words, $\Pi^\mathrm{U^<}_\mathrm{U^>}$, which is the only cross-scale flux
in incompressible hydrodynamics, is completely suppressed here and the cross-scale energy transfer
transfer from the kinetic energy budget is solely mediated by magnetic tension.

\subsubsection{Large scale energy conversion}
\begin{figure}[htbp]
\centering
\includegraphics{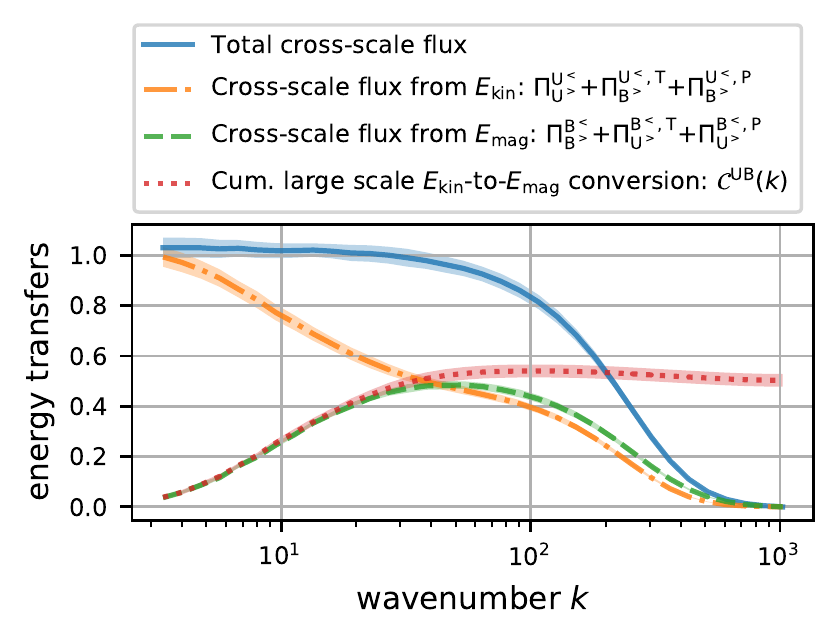}
\caption{Cross-scale energy transfer across $k$ from the
kinetic budget (orange) and magnetic budget (green), and cumulative
energy conversion from kinetic to magnetic energy on scales larger
than $k$ in the stationary regime.
}
\label{fig:cross-vs-k}
\end{figure}

While cross-scale fluxes allow for intra- (via advection) and inter-budget (via magnetic
tension and pressure) transfers, only the latter contributes to a conversion of energy.
Figure~\ref{fig:cross-vs-k} illustrates the net cross-scale fluxes versus scale 
in the stationary regime along with the cumulative large scale kinetic to magnetic
energy conversion.
The cumulative large scale conversion refers to the net energy transfer between those
two budgets on all scales larger than the reference scale $k$,
\begin{align}
\mathcal{C}^\mathrm{UB}(k) = \sum_{Q,K \leq k} \T{UBT}(Q,K) +  \T{UBP}(Q,K)\;,
\end{align}
where the magnetic pressure contribution is negligible in the simulation
presented here.
The cumulative energy conversion tightly follows the the cross-scale flux
from the kinetic energy budget.
On the largest scales ($k\approx4$) it is negligible.
Here, the cross-scale flux is dominated from the kinetic energy budget as expected 
in a situation with a large scale mechanical driving.
From the large to intermediate scales ($k\approx30$) the contribution
continuously grows while the kinetic energy cross-scale flux contribution decreases.
Eventually, the kinetic and magnetic cross-scale
fluxes become approximately the same strength.
Similarly, the cumulative energy conversion reaches a constant value.
This implies that no significant net energy conversion occurs on
intermediate and small scales and is in agreement with \citet{Bian2019},
who show that mean field line stretching is a predominantly
large-scale process.

\section{Summary, Discussion \& Conclusions}
\label{sec:summary}

Motivated by an apparent discrepancy between kinetic and magnetic energy spectra scalings
measured in simulations \citep{ Haugen2004,Moll2011,Teaca2011,Eyink2013,Porter2015,Grete2017a,Bian2019}
and observations of the solar wind \citep{Boldyrev2011} with expectations derived from analytic theory \citep{Galtier2016,Beresnyak2019}, we presented shell-to-shell energy transfer analysis of an implicit large eddy simulation of approximately isothermal, subsonic, super-Alfv\'enic MHD turbulence with vanishing background magnetic field, cross-helicity, and magnetic helicity. In the context of this analysis, we find that magnetic tension suppresses the kinetic energy cascade resulting in a spectrum that is shallower than predicted in
various theories, e.g.,
$k^{-3/2}$ \citep{Iroshnikov1964,Kraichnan1965}.
Overall, the results presented here demonstrate that the energy flux across scales is dominated by magnetic tension, and
similarly the scale local energy balance in the stationary turbulence regime is maintained by a constant energy transfer between the kinetic and magnetic reservoir mediated by magnetic tension.

The simulations on which the results are based, are necessarily limited.
While a clear signature of an extended range with a scaling close to
$k^{-4/3}$ is observed in the kinetic energy power spectrum, no
such range is observed in the magnetic energy power spectrum (see
Figure~\ref{fig:slope-comp}).
We attribute this to a combination of the simulation setup as well as
a limited dynamical range.
More specifically, the mechanical energy injection on the largest scales
provides a barrier for the large scale magnetic field growth in the absence of a
significant (external) mean field.
As a result, the magnetic field is strongest on intermediate scales and
gets weaker towards larger scales.
Similarly, in the limit of large Reynolds numbers we expect the ratio of 
$\Emag(k)/\Ekin(k)$ to grow from the smallest (non-dissipative) scales
towards larger scales until the growth is inhibited by the forcing
acting on the largest scales.
This also explains why extended scaling ranges are regularly observed 
in reduced MHD simulations or in simulation with a significant mean field
(potentially stronger than the velocity field on the forcing scales) where
it, figuratively, provides a large scale anchor, see \citep{Beresnyak2019}.

In this study, we focus on simulations with magnetic Prandtl numbers
of $\Pm \simeq 1$ -- that is, calculations where the kinetic viscosity
and magnetic diffusivity are approximately the same.  We note that the
results presented here appear to be generally independent of numerical
method in the $\Pm \approx 1$ regime.  As shown in
Figure~\ref{fig:slope-comp}, the scaling in the kinetic energy spectrum
has been observed in pseudospectral, finite difference, and finite
volume simulations, and with or without explicit (hyper)dissipative
terms.  While the relative behavior on the smallest scales will depend
on $\Pm$, overall it is expected that for $\Rm > \Re$ (i.e., $\Pm >
1$), where magnetic diffusivity is very low compared to kinetic
viscosity, magnetic energy will be amplified above the kinetic energy
on all scales smaller than the energy injection scale, with the
opposite effect happening in the $\Pm < 1$ regime
\citep{Brandenburg2014a}.  While shell models suggest that the the
magnetic field will continue to show the behavior we have observed in
the $\Pm \gg 1$ regime, at $\Pm \ll 1$ (i.e., when the magnetic
diffusion rate is high) it is likely that there will be very little
magnetic power at small scales, although the precise details will
likely depend on the nature of the turbulent driving.  Given that a
wide range of magnetic Prandtl numbers are relevant in both
terrestrial and astrophysical systems, further work exploring a wider
range of $\Pm$ is well-motivated.

A further complexity arises when we consider variations in the 
plasma regime.  We are modeling a plasma using the ideal MHD
approximation -- i.e., assuming that particles are highly
collisional, that the Debye length and electron and ion gyroradii are
small, and that the inverse of the electron and
ion cyclotron frequencies are short compared to the spatial and
temporal scales of interest.  As these assumptions are relaxed -- for
example, if the plasma is assumed to be weakly collisional and thus
viscosity and resistivity becomes significantly anisotropic -- this may
impact the results we have observed.  Such regimes are important for
both terrestrial and astrophysical systems, and while they are beyond
the scope of our current efforts they are worthy of consideration.
This may require a substantially different numerical approach, however.
While some deviations from the
ideal MHD regime can be explored with extensions of the MHD
approximation \citep[e.g., adding anisotropic terms as per the Braginskii
approximation,][]{Braginskii65}, it is likely that a kinetic or hybrid fluid/kinetic
approximation will be required for some physical regimes.

The key finding of this work is that magnetic tension acts to suppress cross-scale
kinetic energy transfer, resulting in a kinetic spectrum with a slope $k^{-4/3}$,
in contrast with \emph{theoretical} expectations regarding incompressible hydrodynamic turbulence.
Such a suppression of cross-scale kinetic energy transfer is also observed in \emph{simulations}
of hydrodynamics turbulence, where the ``bottleneck effect'' (a pileup of energy on the smallest scales) results in shallower than $k^{-5/3}$ scaling in the kinetic energy spectrum in hydrodynamic turbulence
\citep{Falkovich1994,Schmidt2006353, Frisch2008,Donzis2010,Kuechler2019,Agrawal2020}.
Recently, \citet{Gong2020} also attributed the hydrodynamic bottleneck effect
to the shallow kinetic energy spectra they observe in their MHD simulations.

The results presented here demonstrate that, contrary to \citet{Gong2020}, the physical mechanism for the shallow
slope of the kinetic energy spectrum is \emph{fundamentally} different between hydrodynamics and magnetohydrodynamics due to magnetic tension (which is naturally absent 
in hydrodynamics) causing a suppression of the kinetic cascade cross-scale flux.
In addition, the results presented here suggest that the kinetic cascade is practically absent instead of being
decoupled (from a magnetic cascade), as was recently suggested by \citet{Bian2019}.
While we still observe a significant energy flux in the magnetic energy cascade,
the balance in the kinetic energy budget is maintained by magnetic tension.
Thus, both energy budgets remain coupled through dynamically significant energy fluxes.
Note, compared to the vastly extended dynamical range in
\citet{Bian2019} (which comes from 
the use of higher-order hyperdisspative terms), the dynamical range in the
simulation presented here is rather limited.
Future simulations with a larger dynamical range will help to address this question.

With these caveats in mind, the results presented here have a number of implications.
First, they motivate a reevaluation of MHD turbulence theories that commonly are only concerned with the total (kinetic and magnetic) energy spectrum and energy flux.
In particular, the results presented here suggest that flux-based models should differentiate between the
intra- and inter-budget cross-scale fluxes, and consider energy budgets
separately. We note that the scaling of the total energy will be dominated  by the magnetic
energy scaling on intermediate scales, 
which is important in the light of MHD turbulence theory on scaling relations.
Second, in the interpretation of observations and their derived spectra special
care is required in inferring properties from one spectra to the other, as
we see no indication that that kinetic and magnetic energy spectra follow
the same scaling laws, \citep[cf., also][]{Boldyrev2011}.
Third, subgrid-scale modeling in the context of large eddy simulations 
\citep{Miesch2015,Grete2016a} may become simpler as, for example,
one can neglect a purely kinetic cross-scale flux.
Finally, we note that in natural systems the effective large scale driving mechanisms, e.g., a galaxy cluster merger \citep{Subramanian2006}, provides an outer scale and limit for the amplification of magnetic fields by the fluctuation dynamo.

Finally, we note that our results should also be interpreted with care
and not be overgeneralized.
As mentioned in the Introduction, the configuration space of MHD turbulence
is vast and the results presented here cover only a single point.
Additional data from (even larger-scale) simulations, observations, and 
experiments is required in order to get a complete picture of MHD turbulence.

\acknowledgments
The authors thank Jim Stone and Ellen Zweibel for useful discussions.
PG and BWO acknowledge funding by NASA Astrophysics Theory Program
grant \#NNX15AP39G.
Sandia National Laboratories is a multimission laboratory managed and operated by 
National Technology and Engineering Solutions of Sandia LLC, a wholly owned 
subsidiary of Honeywell International Inc., for the U.S. Department of Energy's 
National Nuclear Security Administration under contract DE-NA0003525.
The views expressed in the article do not necessarily represent the views of the 
U.S. Department of Energy or the United States Government.
SAND Number: SAND2020-9337 O.
BWO acknowledges additional funding by NSF grants \#1514700,
AST-1517908 and AST-1908109, and NASA ATP grant 80NSSC18K1105.

The simulations and analysis were run on the NASA Pleiades supercomputer through allocation SMD-16-7720, 
on TACC's Stampede2 supercomputer as part of the Extreme Science and Engineering Discovery Environment \citep[XSEDE][]{XSEDE}, which is supported by National Science Foundation grant number ACI-1548562, through allocation \#TG-AST090040,
and on TACC's Frontera supercomputer through LRAC allocation \#AST20004.

The software below is developed by a large number of independent
researchers from numerous institutions around the world. Their
commitment to open science has helped make this work possible.
\software{
  \kathena \citep{kathena}, a performance portable version of
  \athenapp \citep{Stone2019} using \kokkos \citep{Edwards2014}.
  \texttt{Matplotlib} \citep{matplotlib}.
  \texttt{NumPy} \citep{numpy}.
  \texttt{mpi4py} \citep{mpi4py}.
  \texttt{mpi4py-fft} \citep{mpi4py-fft}.
}


\end{document}